\documentclass[conference]{IEEEtran}
\IEEEoverridecommandlockouts
\usepackage{cite}
\usepackage{amsmath,amssymb,amsfonts}
\usepackage{algorithmic}
\usepackage{graphicx}
\usepackage{textcomp}
\usepackage{xcolor}
\def\BibTeX{{\rm B\kern-.05em{\sc i\kern-.025em b}\kern-.08em
    T\kern-.1667em\lower.7ex\hbox{E}\kern-.125emX}}

\usepackage{hyperref}

\begin{document}

\title{GestureKeeper: Gesture Recognition for Controlling Devices in IoT Environments
\thanks{\IEEEauthorrefmark{5}Contact author: Maria Papadopouli  (mgp@ics.forth.gr)}
}

\author{\IEEEauthorblockN{
Vasileios Sideridis\IEEEauthorrefmark{1},
Andrew Zacharakis\IEEEauthorrefmark{1}\IEEEauthorrefmark{2},
George Tzagkarakis\IEEEauthorrefmark{1}, and
Maria Papadopouli\IEEEauthorrefmark{1}\IEEEauthorrefmark{2} \\
\IEEEauthorblockA{\IEEEauthorrefmark{1}Institute of Computer Science, Foundation for Research and Technology-Hellas, Heraklion, Greece\\
\IEEEauthorrefmark{2}Department of Computer Science, University of Crete, Heraklion, Greece\\
}}}
\maketitle

\begin{abstract}
This paper introduces and evaluates the GestureKeeper, a robust hand-gesture recognition system based on a wearable inertial measurements unit (IMU). The identification of the time windows where the gestures occur, without relying on an explicit user action or a special gesture marker, is a very challenging task. To address this problem, GestureKeeper identifies the start of a gesture by exploiting the underlying dynamics of the associated time series using a recurrence quantification analysis (RQA). RQA is a powerful method for nonlinear time-series analysis, which enables the detection of critical transitions in the system's dynamical behavior. Most importantly, it does not make any assumption about the underlying distribution or model that governs the data. Having estimated the gesture window, a support vector machine is employed to recognize the specific gesture. Our proposed method is evaluated by
means of a small-scale pilot study at FORTH and demonstrated that GestureKeeper can identify correctly the start of a gesture with a 87\% mean balanced accuracy and classify correctly the specific hand-gesture with a mean accuracy of over 96\%. To the best of our knowledge, GestureKeeper is the first automatic hand-gesture identification system based only on accelerometer. The performance analysis reveals the predictive power of the features and the system's robustness in the presence of additive noise. We also performed a sensitivity analysis to examine the impact of various parameters and a comparative analysis of different classifiers (SVM, random forests). Most importantly, the system can be extended to incorporate a large dictionary of gestures and operate without further calibration for a new user.
\end{abstract}


\section{Introduction}
\label{sec:intro}

In general, hand-gesture recognition systems can be classified in two categories, according to the type of sensors they employ, namely, the camera- and the wearable-based ones. The camera-based systems can achieve high recognition accuracy, but at a relatively high computational cost~\cite{[vision2]}. The performance of these systems is sensitive in the background, light conditions, and room geometry, and constrained by the field of view of the camera. On the other hand, sensor-based systems, worn on the wrist, leg, arm, chest, ankle, head and/or waist, employ accelerometers, gyroscopes and magnetometers, barometers, body sensor networks~\cite{[BSN_paper1]}, electromyography sensors~\cite{[EMG_paper]}, or even sound sensors~\cite{[Math_paper3]}. They have relatively small cost, are not particularly sensitive to the environmental conditions (e.g. light or geometry conditions), and can function in indoor and outdoor spaces. Although, in general, wearable sensors are energy-constrained, due to the rapid advances of the micro-electromechanical technologies in reducing their size and enhancing their energy efficiency, they have gained a lot of attention in hand-gesture recognition (HGR), human activity recognition (HAR)~\cite{[RQA1],[RQA3],[HAR1],[HAR2],[HAR3]} and even in human writing recognition~\cite{[HWR_paper1]}. These systems collect data from the on-board sensors and apply either machine learning algorithms~\cite{[ML_paper1],[ML_paper3],[ML_paper5],[ML_paper6],[ML_paper7],[ML_paper8]}, mathematical models~\cite{[Math_paper3],[Math_paper1],[Math_paper2]}, fuzzy control techniques~\cite{[fuzzy_paper1]} or simple threshold-based algorithms~\cite{[Threshold_paper2]} to recognize the user's gestures. Inertial measurement units are extremely useful and commonly-used for orientation/heading estimation~\cite{[orientation_paper1],[orientation_paper2]}. The prevalence of accelerometers, gyroscopes and magnetometers in smart-phones for estimating their orientation has enabled various interesting applications in the gesture recognition domain. 


This paper introduces and evaluates the GestureKeeper, an innovative robust hand-gesture identification and recognition system based on a wearable inertial measurements unit (IMU). In the context of daily activities, the user can control an appliance by performing a specific gesture. The identification of starting and ending points of the time windows, where the gestures occur, without relying on an explicit user action, as in~\cite{[ML_paper1]}, or a special gesture marker, as in~\cite{[StartStopGesture]}, is challenging. To address this problem, GestureKeeper identifies the start of a gesture by exploiting the underlying dynamics of the associated time series using recurrence quantification analysis (RQA). RQA is a powerful method for nonlinear time-series analysis, which enables the detection of critical transitions in the system's dynamics (e.g. deterministic, stochastic). Most importantly, it does {\em not} make any assumption about the underlying distribution or model that governs the data. Moreover, it can be used even for relatively small and non-stationary datasets.
More specifically, our proposed method capitalizes on the efficiency of RQA to extract the underlying dynamics of a recorded sensor data stream by mapping the associated time series in a higher-dimensional phase space of trajectories. A major advantage of RQA is its \textit{fully self-tuned nature}, in the sense that no prior parameter fine-tuning is required in a manual fashion. 

To the best of our knowledge, GestureKeeper is the {\em first automatic hand-gesture identification system based only on accelerometers}. Furthermore, it can recognize accurately a dictionary of 12 hand-gestures by applying support vector machines (SVM) on a hybrid set of statistical and sample-type features. The evaluation was performed in a small-scale pilot study at FORTH. This paper demonstrates that GestureKeeper can recognize gestures from our 12-hand-gesture dictionary with a mean accuracy of about  96\%. The analysis also reveals the predictive power of the features and the system's robustness in the presence of additive noise. We performed a sensitivity analysis to examine the impact of various parameters. Finally, to comparatively assess the performance of SVM, we also applied random forests. SVM still achieves a higher accuracy.
The rest of the paper is organized as follows: Section~\ref{sec:sys} overviews our system, its architecture, and the two main sub-systems, while Section~\ref{sec:concl} summarizes our main conclusions and future research directions.

\section{GestureKeeper System Design} \label{sec:sys}
GestureKeeper consists of a wearable sensor and a server. The wearable sensor sends periodically its collected measurements to the server, while the server performs the gesture identification and recognition. Our dictionary consists of 12 gestures. Their names, short descriptions, and trajectories in space are shown below.

\begin{itemize}
\setlength{\itemsep}{-0.2\baselineskip}
  \item {\makebox[1.2cm]{\textbf{Up:}\hfill} Vertical movement towards the ceiling}
  \item {\makebox[1.2cm]{\textbf{Down:}\hfill} Vertical movement towards the ground}
  \item {\makebox[1.2cm]{\textbf{Left:}\hfill} Horizontal movement to the left}
  \item {\makebox[1.2cm]{\textbf{Right:}\hfill} Horizontal movement to the right}
  \item {\makebox[1.2cm]{\textbf{CW:}\hfill} Clock-wise rotational movement}
  \item {\makebox[1.2cm]{\textbf{CCW:}\hfill} Counter Clock-wise rotational movement}
  \item {\makebox[1.2cm]{\textbf{Z:}\hfill} Z trajectory starting from above}
  \item {\makebox[1.2cm]{\textbf{AZ:}\hfill} Mirror Z trajectory starting from below}
  \item {\makebox[1.2cm]{\textbf{S:}\hfill} Wave trajectory to the right}
  \item {\makebox[1.2cm]{\textbf{AS:}\hfill} Wave trajectory to the left}
  \item {\makebox[1.2cm]{\textbf{Push:}\hfill} Horizontal movement away from the body}
  \item {\makebox[1.2cm]{\textbf{Pull:}\hfill} Horizontal movement towards the body}
\end{itemize}

\begin{figure}[ht]
	\includegraphics[scale=0.45]{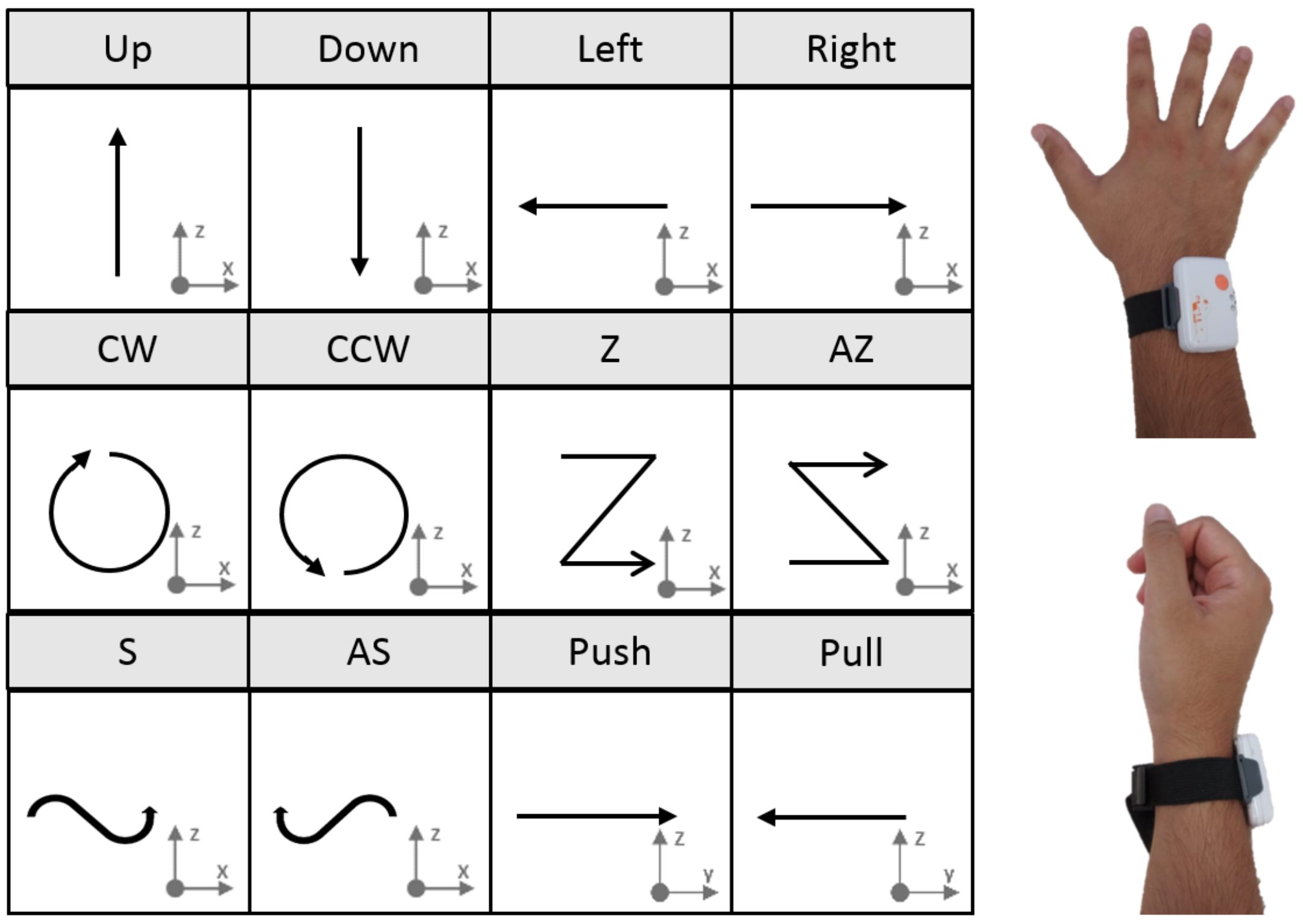}
	\centering
	\caption{Dictionary of gestures and hand placement.}
	\label{gestures}
\end{figure} 

{\bf Experiments and Data Collection.} We employed the shimmer3~\cite{[shimmer3]}, a wearable device equipped with 3-axis accelerometer, gyroscope, and magnetometer. We first calibrated the sensors and then configured each of them according to the needs of our implementation. For example, due to the relatively low expected acceleration, we chose a relatively small detection range for the accelerometer increasing this way its resolution. The final sampling frequency was set to 50 Hz, which is considered sufficient (given our observations from a number of fast steep movement experiments). An additional increase of the frequency will only increase the power consumption of the device and the measurement size without, however, enhancing the amount of information about the user's movement. Finally, the data streaming and logging applications are responsible to collect the data during the experiments.

For the recognition sub-system, we performed a small field study with 15 subjects (9 female, 6 male). Each subject repeated a number of gestures (from the predefined dictionary of gestures), collecting in total 900 different repetitions. For each repetition, 63 statistical features and 30 acceleration-type features, which represent values of acceleration in the x, y and z axes, were extracted. This dataset had the isolated periods during which a gesture was performed. For the identification sub-system, a {\em new} dataset was produced, containing gestures as well as activities of daily living (ADL). The new dataset has a total duration of 3 hours and 45 minutes, with measurements collected from 4 new subjects (2 female, 2 male).   
\subsection{Gesture Identification} \label{sec:sys:identification}
We first focus on the problem of identifying the start of a gesture in a recorded data stream. We speculate that different gestures are characterized by distinct dynamics of the associated time series. This motivated the use of the RQA, which enables the detection of transitions in the dynamical behavior (e.g. deterministic, chaotic, etc.) of the observed system. A major advantage of RQA is its \textit{fully self-tuned nature}, in the sense that no prior parameter fine-tuning is required in a manual fashion. More specifically, a recurrence plot (RP) is derived first, which depicts those times at which a state of a dynamical system recurs, thus revealing all the times when the \textit{phase space trajectory} of the dynamical system visits roughly the same area in the phase space. To this end, RPs enable the investigation of an $m$-dimensional phase space trajectory through a two-dimensional representation of its recurrences. Such recurrence of a state occurring at time $i$, at a different time $j$ is represented within a two-dimensional square matrix with ones (recurrence) and zeros (non-recurrence), where both axes are time axes. Given a time series of length $N$, $\{r_i\}_{i=1}^N$, a phase space
trajectory can be reconstructed via time-delay embedding,
\begin{equation}
{\bf x}_i = [r_i,\,r_{i+\tau},\ldots,r_{i+(m-1)\tau}]\ ,\quad i=1,
\ldots,N_s\ ,
\label{eq:TDEmb}
\end{equation}
where $m$ is the embedding dimension, $\tau$ is the delay, and
$N_s=N-(m-1)\tau$ is the number of states.
Having constructed a phase space representation, an RP is defined
as follows,
\begin{equation}
{\bf R}_{i,j} = \Theta\left(\varepsilon - \|{\bf x}_i - {\bf x}_j\|_p\right)\ ,
\quad i,\,j = 1,\ldots,N_s\ ,
\label{eq:RPdef}
\end{equation}
where ${\bf x}_i$, ${\bf x}_j\in\mathbb{R}^m$ are the states,
$\varepsilon$ is a threshold, $\|\cdot\|_p$ denotes a
general $\ell_p$ norm, and $\Theta(\cdot)$ is the Heaviside step
function, whose discrete form is defined by
\begin{equation}
\Theta(n) = \begin{cases}
1, & \text{if $n\geq 0$} \\
0, & \text{if $n<0$}
\end{cases} \ ,\ n\in\mathbb{R}\ .
\label{eq:HeaviFun}
\end{equation}

Typically, several linear
(and/or curvilinear) structures appear in RPs, which provide hints
about the time evolution of the high-dimensional phase space
trajectories. Besides, a major advantage of RPs is that they can
also be applied to rather \textit{short} and even
\textit{non-stationary} data. 
The visual interpretation of RPs, which is often difficult and subjective, is enhanced by means of several numerical measures for the quantification of the structure and complexity of RPs~\cite{[RQA]}. These quantification measures provide a global picture of the underlying dynamical behavior during
the entire period covered by the recorded data. The temporal evolution of RQA measures and the subsequent detection of transient dynamics are enabled for each recorded sensor stream by employing a windowed version of RQA. Doing so, the corresponding quantification measures are computed in small windows, which are then merged to form our feature matrix. Furthermore, it is noted that the length of the sliding window yields a compromise between resolving small-scale local fluctuations and detecting more global recurrence structures.

The gesture identification of the GestureKeeper employs two RQA metrics, namely, the recurrence rate (RR) and the transitivity (TRA) (see~\cite{marwan2007recurrence} for the definitions), obtained from the y-axis acceleration data, in order to form the feature matrix\footnote{We initially employed an extended RQA measures set but observed that the aforementioned two measures are sufficient to identify the gesture time-windows.}. 

{\bf Estimation of Embedding Parameters.} In our implementation,
the optimal time delay $\tau$ is estimated as the first minimum of
the average mutual information (AMI) function~\cite{FraSwi86}.
Concerning the embedding dimension $m$, a minimal sufficient value
is estimated using the method of false nearest neighbours
(FNN)~\cite{KBAb92}. 
Furthermore, the Euclidean norm is used as our selected distance metric
for the construction of the RP, while a rule-of-thumb is currently
used to set the threshold $\epsilon=0.2\sqrt{m}$. The window length refers to the size of the signal in which RQA is performed each time before being shifted by a step size to the next values. We selected a window size of 125 samples, which represents approximately 2.5 seconds of information, sufficient for capturing even the longest of the dictionary's gestures. The step was set equal to 25 samples (i.e., 80\% overlap between consecutive windows). 
By applying the above criteria on our data, the estimated embedding parameters are equal to $m=4$ and $\tau=1$. Although the empirical rule for selecting $\epsilon$ yields $\epsilon = 0.4$, a higher accuracy was observed for $\epsilon = 0.1$. 


An SVM classifier based on these two features was then employed to classify the data in two classes, namely, gestures and ADL, thus distinguishing the gestures from the rest of the hand movements.

\subsection{Gesture Recognition} \label{sec:sys:recognition}
The gesture recognition is based on an SVM classifier using the radial kernel\footnote{Radial kernel was shown in our sensitivity analysis to have the best accuracy among the polynomial, linear, and sigmoid.}. For the classification, we employed two types of features, namely the {\em statistical features} and {\em samples of the acceleration signal}. The statistical features include the mean, median, root mean square (RMS), standard deviation, variance, skewness and kurtosis, of the 3D acceleration, angular velocity, and magnetism time series provided by the sensor. The sample based features are formed by a re-sampling process of the x, y and z axis acceleration\footnote{The original acceleration signal consists of 3 time series, one for each axis, of unknown length (since it depends on the particular gesture and the user who performed the movement).}. After re-sampling, the new signal is composed of a fixed number of samples for each acceleration time series. The final set of features includes the statistical ones (introduced earlier) along with this number of samples of the re-sampled acceleration signal for each of the x-, y-, and z-axis time series.
\subsection{Performance Analysis}

\textit{Gesture Identification}: GestureKeeper first utilizes RQA in order to extract features for the identification sub-system, thereafter uses an SVM model, trained with the aforementioned features, for identifying correctly the windows that contain gestures. Note that the ADL and gesture classes are highly unbalanced: only 0.5\% of the data belong to the class ``gestures'' and the 99.5\% is ADL. To analyze the performance of the identification process, we trained the SVM classifier (with a polynomial kernel of degree = 3, coefficient = 2, cost = 3, and gamma = 0.95\footnote{We also examined various classifiers using different kernels, such as linear, sigmoid, radial, and parameters values. The reported results were obtained using the above values.}) using all subjects except one, which was then used for testing. Given that the ADL and gesture classes are highly unbalanced, we randomly selected a subset of the ADL class of equal size as the gesture one. We performed the training and testing on this dataset. This process was repeated for 100 iterations. We reported the mean accuracy for each subject of the testing. The accuracy for each subject varies from 76.9\% to 91.1\% (with a mean accuracy of 87.21\%).

\begin{table}[ht]
\scriptsize
\caption{List of all the parameters that were used in GestureKeeper}
\label{table1}
\begin{tabular}{llllll}
\multicolumn{2}{l}{\textbf{RQA}} & \multicolumn{2}{l}{\textbf{SVM Identification}} & \multicolumn{2}{l}{\textbf{SVM Recognition}} \\
Parameters         & Value       & Parameters                & Value               & Parameters              & Value              \\
Disntance Metric   & Eucl. Norm  & Kernel                    & polyn.              & Kernel                  & radial             \\
Window size        & 125         & Gamma ($\gamma$)                 & 0.95                & Gamma ($\gamma$)               & 0.005              \\
Window step        & 25          & Cost (c)                  & 3                   & Cost (c)                & 1                  \\
Delay ($\tau$)          & 1           & Degree                    & 3                   &                         &                    \\
Dimensions (m)     & 4           & Coefficient               & 2                   &                         &                    \\
Threshold ($\epsilon$)      & 0.1         &                           &                     &                         &                   
\end{tabular}
\end{table}

\begin{figure}[ht]
	\includegraphics[scale=0.41]{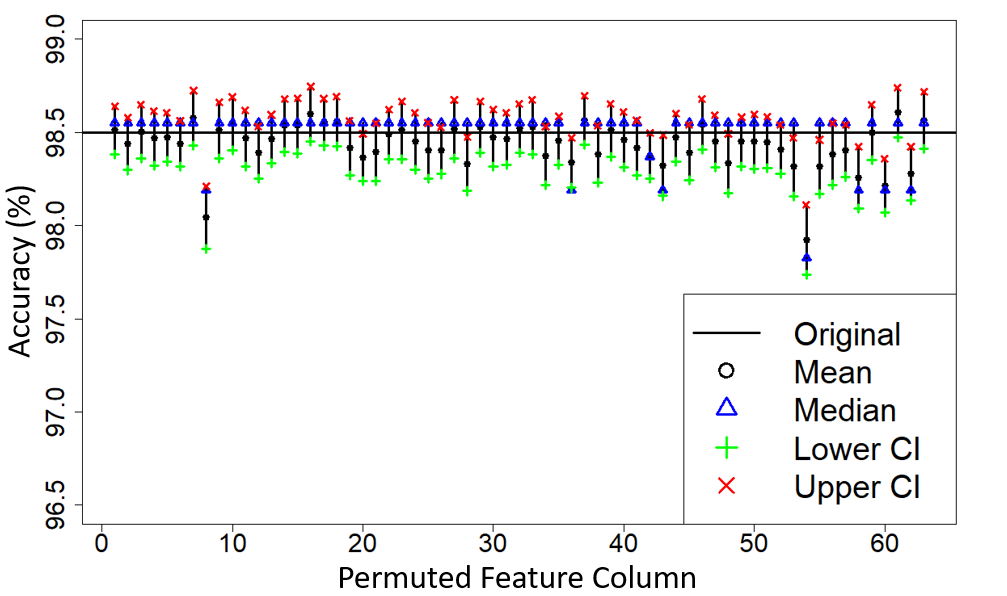}
	\centering
	\caption{Mean accuracy under a single-feature permutation (indicated in the x-axis). Features are based on simple statistics of data collected from accelerometer, gyroscope, and magnetometer from each of the three dimensions. Original line: no feature has been permuted.}
	\label{statistical_Permutations1C}
\end{figure}

\textit{Gesture Recognition}: As mentioned earlier in Sec. \ref{sec:sys:recognition}, an SVM model with different parameters is employed for the final classification of the gestures in one of the 12 in total classes of our dictionary. To assess the predictive power of the statistical features, we performed the following process: First, we permuted a feature, keeping the values of the remaining features of the dataset fixed. Then, the model is trained and tested with this dataset. The mean accuracy is based on 100 repetitions. We then repeated the same process, selecting each time to permute a different feature from the original dataset. The multiclass classification model employs the ``one-against-one'' approach, in which $12\cdot(12-1)/2$ binary classifiers are trained and the appropriate class is finally selected by a voting scheme. The mean accuracy for the different permuted features is shown in Fig.~\ref{statistical_Permutations1C}. The horizontal line indicates the mean accuracy for the original dataset without any feature permutation. The larger the decrease in the accuracy, the more significant the information that the corresponding feature provides in gesture recognition. It appears that the mean and skewness of the z-axis angular velocity are the most significant ones. 

\begin{figure}[htb]
	\includegraphics[scale=0.41]{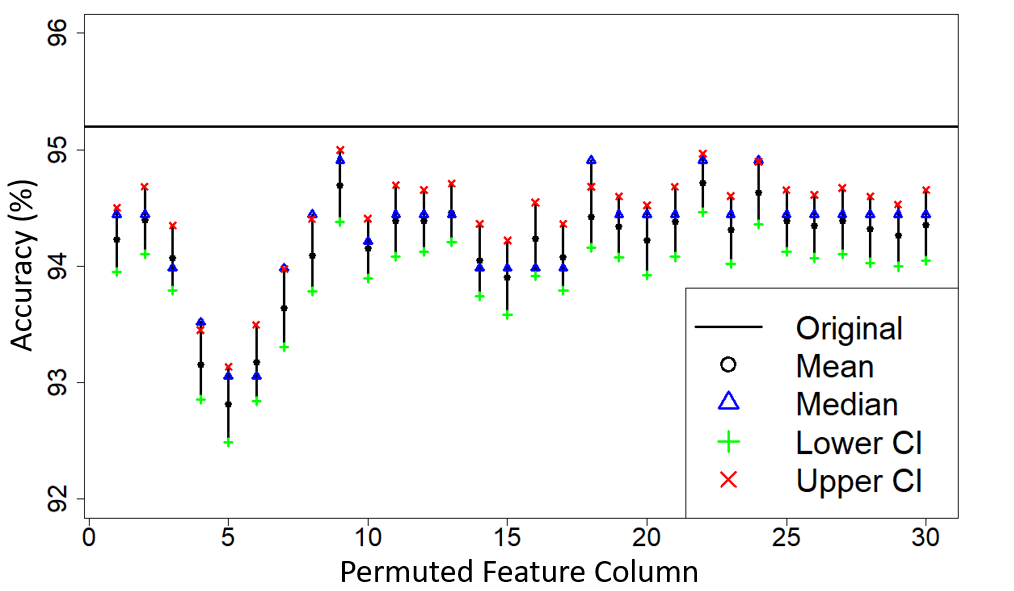}
	\centering
	\caption{Mean accuracy under a single-feature permutation (indicated in the x-axis). The first ten features correspond to a re-sampling of the acceleration data in the x-axis, while the 11th-20th (21th-30th) features correspond to y-axis (z-axis), respectively. Original line: no feature has been permuted.}
	\label{samples_Permutations1C}
\end{figure}

\begin{figure}[ht]
	\includegraphics[scale=0.25]{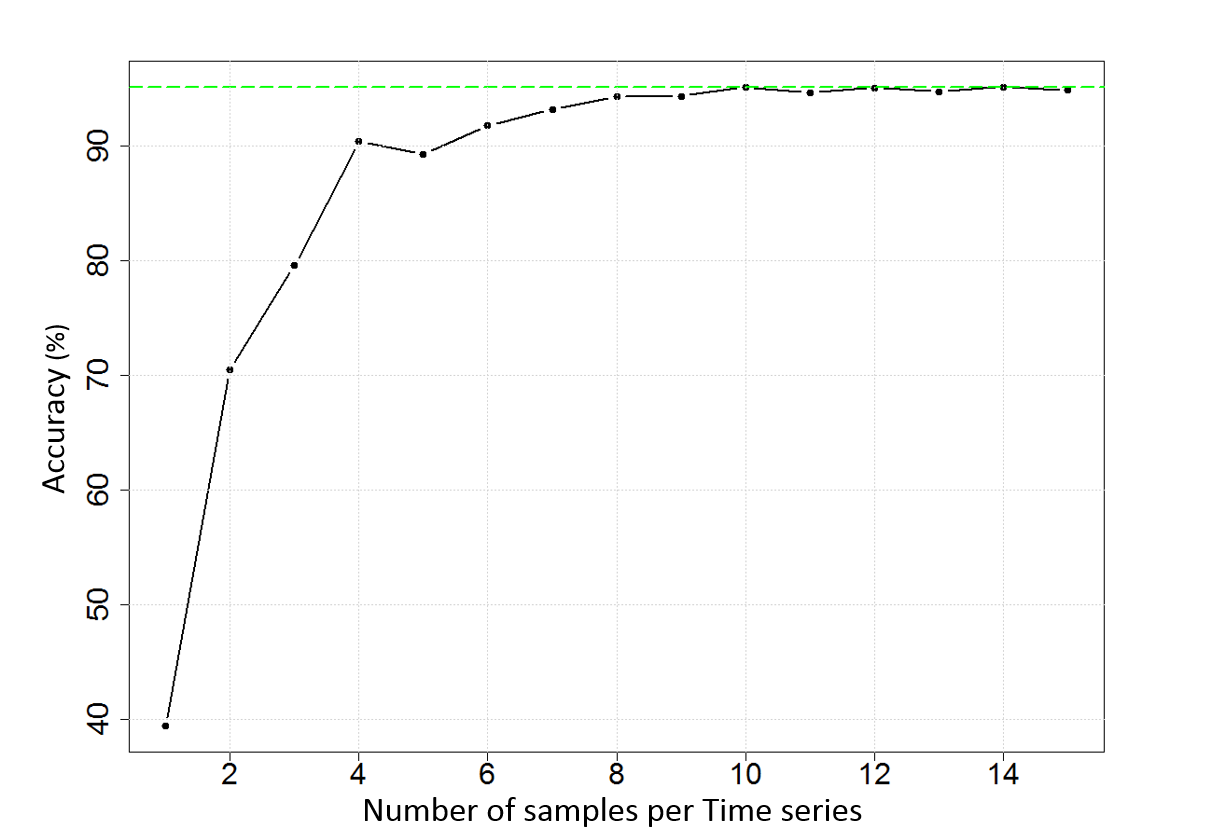}
	\centering
	\caption{Mean accuracy vs. number of samples per time series.}
	\label{acc}
\end{figure}

A similar procedure was also applied for the acceleration-sample-based features. We examined a different number of samples for each time series (in the range of 4-12). In the case of ten samples per time series, the fourth, fifth, sixth value of the re-sampled x-axis acceleration have a significant impact on accuracy, while the remaining features exhibit a similar predictive power for the gesture recognition (as shown in Fig.~\ref{samples_Permutations1C}). However, unlike the statistical features, all the acceleration-sample-based ones have a substantial contribution on the accuracy (as the accuracy using the original dataset is greater from the one using the permuted ones). Note that the first 10 samples that refer to acceleration in x axis carry more information than the others. Specifically, the most important feature tends to be the middle value of the x-axis signal (sample 5). Fig.~\ref{acc} shows the accuracy of the gesture classification, when only the sample features are used, as a function of the sample size. The impact on the accuracy is prominent for a sample size of 5 or less.

Our gesture recognition classifier employs in total 73 features, namely the 43 most significant statistical features and 10 samples-based features from each acceleration time series. The above analysis was performed using the default SVM hyper-parameters, namely, cost $c=1$ and $\gamma=1/$number of features.

{\bf Experiments and SVM Tuning.} We train the gesture recognition classifier using the data collected from the 14 out of 15 subjects for training and the other one's for testing. We performed tests for all the 15 combinations of training and testing partitions. 

\begin{figure}[ht]
	\includegraphics[scale=0.39]{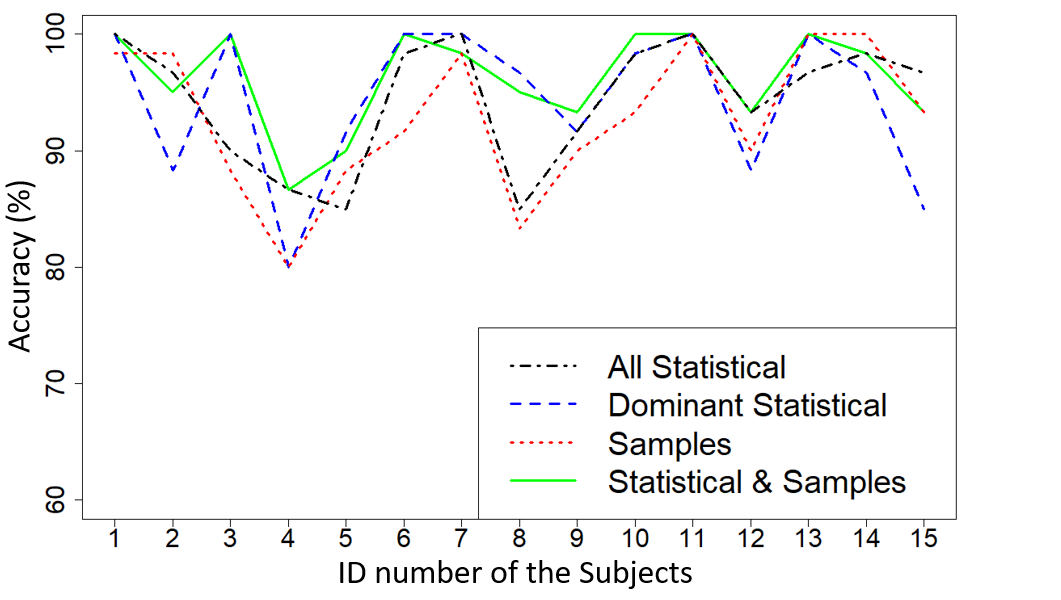}
	\centering
	\caption{Mean accuracy of each subject's data for different datasets.}
	\label{accuracy_tests}
\end{figure}

The hyper-parameters that we tuned are the cost (c) and gamma ($\gamma$) values. The cost represents the weight for penalizing the "soft margin". Consequently, a large cost value penalizes the SVM for data points within the margin or on the wrong side of the dividing hyper-plane. For this reason, for large $c$, SVM will try to find more complex hyper-planes, and possibly smaller margins that leave less data points in the wrong side, but is also more prone to over-fitting. In contrast, for low cost values, the margins of SVM will be larger and the hyper-plane less complex, making it more robust but also less accurate. This motivates the need for carefully addressing the tradeoff of accuracy and robustness. The gamma parameter refers to the kernel and depends on the number of features that each specific implementation has. Intuitively, gamma "controls" the number of SVM's support vectors and by extension the sensitivity of the decision boundary (hyper-plane), which will affect whether or not some data points near the margins will be ignored. A sensitivity analysis with different cost and gamma values reports the best performance for $c = 1, \gamma = 0.005$. Fig.~\ref{accuracy_tests} shows the accuracy of each subject's data as testing set for the following datasets, namely all statistical features, only significant statistical ones, only samples, both significant statistical features and the samples (proposed dataset), with a mean accuracy of 94.44\%, 94.44\%, 92.88\%, and 96.22\%, respectively. The best performance is obtained when all the significant statistical features and samples-based ones are employed. The presence of all the parameters that were used in this paper is considered essential for any further experimental evaluation. Therefore Table \ref{table1} shows the parameters for the RQA and the SVM model of the first and second sub-system, respectively.

To comparatively evaluate the recognition sub-system, we also developed classifiers based on random forests. We evaluated them using the (first) dataset (that was also used for the recognition sub-system based on SVM). After training and tuning\footnote{The sensitivity analysis for the random forests reported best accuracy for 100 trees in the forest and a maximum depth of the tree equal to 10. The minimum number of samples required to split an internal node and the minimum number of samples required to be at a leaf node were left with the default values of 2 and 1 respectively as they did not seem to effect the accuracy.}, random forests reported a best accuracy of 88\% as opposed to 96\% of the proposed method using SVM. 

Finally, we assessed the impact of noise on the accuracy of the gesture recognition sub-system. The mean accuracy was increased to 97.89\%, when each feature vector of the original (clear) training dataset was augmented by a copy of a corrupted one (produced by adding Gaussian noise with standard deviation of 0.5).


\section{Conclusion}  \label{sec:concl}
This paper presents GestureKeeper which employs an accelerometer, gyroscope and magnetometer, from a wearable IMU, to first identify time-windows that contain a gesture, and then, recognize which specific gesture it is. GestureKeeper uses features based on statistical properties and acceleration samples. It can accurately recognize gestures from our 12-hand-gesture dictionary, exhibiting its best performance when the combination of features are used (e.g., about 96\% mean accuracy). With the noise addition and feature selection, the mean accuracy is increased to over 97\%. It is modular and can be extended to recognize a larger gesture dictionary size. 
The pilot field study was performed in a relatively controlled small-scale environment. We plan to extend the evaluation under more realistic conditions.
Moreover, it is critical to correctly center the gesture for each of the identified time-windows, as we observed a significant accuracy drop.
We will explore the use of long-short term memory (LSTM) networks and conditional random fields in the gesture recognition  to address these challenges.






\end{document}